\documentclass[3p,times]{elsarticle}

\usepackage{amssymb}

\usepackage[figuresright]{rotating}
\usepackage{multicol}
\usepackage[fleqn]{amsmath}
\usepackage{caption}
\usepackage{graphicx}
\usepackage{subfigure}
\usepackage{setspace}
\begin{document}



 

\title{Spin-orbit coupled transport in a curved quantum wire}


\author{C.~Baldo\corref{Baldo}} 
\cortext[Baldo]{Tel.: +63-2-920-9749}
\ead{cbaldo@nip.upd.edu.ph}
\author{C.~Villagonzalo}

\address{Structure and Dynamics Group, National Institute of Physics, University of the Philippines Diliman, Quezon City, Philippines 1101}


\noindent \textbf{Spin-orbit coupled transport in a curved quantum wire}
\vspace{0.5cm}
\\ 
\noindent Carlos Baldo III$^*$ and Cristine Villagonzalo
\\
\noindent Structure and Dynamics Group, National Institute of Physics, University of the Philippines Diliman, Quezon City, Philippines 1101
\\
$^*$ cbaldo@nip.upd.edu.ph
\vspace{0.5cm}
\\
\textbf{We study the interplay of both Rashba and Dresselhaus spin-orbit couplings (SOCs) and a uniform perpendicular magnetic field $\textbf{B}$ on the transport of a spin-polarized electron along a curved quantum wire. Eigenenergies and eigenfunctions of the system were analytically solved in the presence of both SOCs for a confinement radius $R$. From the transmission coefficients, the spin transport properties  such as spin polarization, probability current density and spin conductance were computed numerically to determine their dependence on the SOCs, $\textbf{B}$ and $R$. We find the condition for $\textbf{B}$ that if it is beyond $R^{1/3}$, no spin reversal will occur. Our results show that for a sufficiently large SOC strength, regardless of its inversion asymmetry origin, the effect of the external magnetic field is reduced. Finding the optimal effective SOC strength is essential in achieving suitable magneto-transport properties for possible spintronic device applications.}




\vspace{0.5cm}

\section{Introduction}
\label{sec:Intro}

Spin-based electronics promises to deliver devices which operate with superior performance (e.g. greater speed, increased memory capacity) as compared to its electric charge-based counterparts. Hence, ongoing efforts have been put forward towards the realization of such  devices which make use of the spin degree of freedom. A large fraction of issues in its development concerns how one can effectively generate and manipulate spin polarized current into a host material without the use of ferromagnetic contacts and/or external micro-magnets \cite{Wolf, Rashba}. These eventually lead to the revived interest in the research of spin-orbit couplings (SOCs) in materials and its potential application in the operation of spin-based devices. \\
\indent The occurrence of SOCs is due to the interaction of the carrier's spin with its orbit. In low-dimensional semiconductor structures, the two common types of SOCs are - the Rashba SOC and the Dresselhaus SOC - which are brought about by the prevailing inversion asymmetry in the material \cite{Lopez}. In semiconductor heterostructures, the Rashba SOC (RSOC) results from the structural inversion asymmetry caused by the confining potential. On the other hand, the Dresselhaus SOC (DSOC) is commonly manifested in III-V semiconductors where there is lack of bulk inversion symmetry \cite{Nakhmedov}. These two types of SOCs may both be present in most semiconductors but one may be stronger than the other.\\
\indent The influence of SOCs on the spin-dependent  transport in mesoscopic systems has been studied in its effects on the transmission of the spin current \cite{Xu, Lu, Xiao}, the possibility of spin switching \cite{Xu, Lu, Jiang, Trushin}, spin conductance \cite{Xu, Jiang} and spin susceptibility \cite{Lopez}. However in mesoscopic loops, several authors also found the dependence of spin transport on the geometry of the system \cite{Trushin, Wang, Wang2, Sil, Liu} particularly on the radius of curvature \cite{Trushin, Wang} and on the magnetic field \cite{Molnar, Sil}. Trushin and Chudnovsky, for instance, reported that the spin polarization can be switched through current density redistribution via non-adiabatic transport in a strongly curved one-dimensional wire wherein RSOC is taken into account \cite{Trushin}.  Just recently, Liu and Xia published their numerical computations on spin-dependent transmission and polarization for which they show that ballistic electron transport in a one-dimensional ring having a square loop configuration with RSOC, in which Zeeman effect is neglected, has higher stability as a spin polarizer compared to a circular loop \cite{Liu}.  Wang and Vasilopoulos, on the other hand, considered both RSOC and DSOC and obtained an analytic expression of conductance which explicitly depends on the ring's radius \cite{Wang}. Note that these studies are not just confined to theoretical investigations as one dimensional quantum wires are already made possible through novel nanofabrication techniques \cite{Yang, Ron}.\\
\indent In this paper, we revisit the curved-one dimensional ballistic wire as used in Ref. \cite{Trushin, Yang2, Bulkagov} to study the transport of a magnetically-polarized electron in the presence of both RSOC and DSOC. Here we seek to determine the effects of $\textbf{B}$, RSOC and DSOC as well as $R$ on the transmission, on the spin orientation of the transmitted spin current and on the conductance. The manuscript is therefore organized as follows: in Section 2 we review the appropriate Hamiltonian to handle a curve geometry with SOC and present our derivation of the corresponding eigenfunctions and eigenvalues, as well as the calculation of the spin polarization, probability spin current density and conductance. In Section 3 we show and discuss the numerical results. Finally, we give the summary of our work in section 4.    

\section{Formalism}
\label{sec:Form}

We investigate the motion of a spin-polarized electron confined in a quasi one-dimensional wire that is under the influence of SOCs and a uniform magnetic field ${\bf B} = B \hat{k}$. Here $\textbf{B}$ is chosen to be perpendicular to the plane of the curved wire. The study of this system's spin dynamics is considered according to three different regions as seen in Fig.~\ref{fig: wire}: a straight input segment, an arc segment of radius $R$ and a straight output segment. In the presence of SOCs, the one-electron Hamiltonian operator has the general form 
\vspace{-0.1cm}

\begin{center}
    \includegraphics[width=0.60\textwidth]{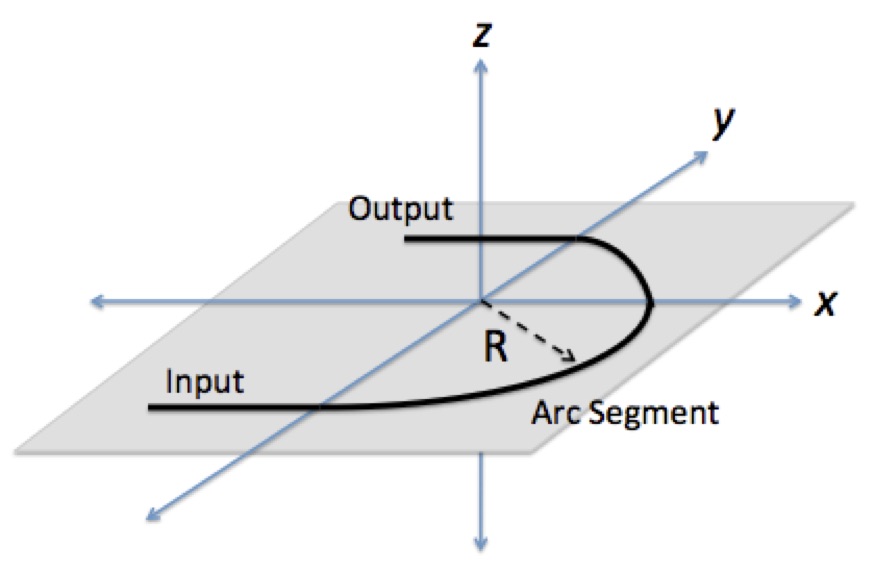}
    \captionof{figure}{Curved quantum wire according to the model of Trushin and Chudnovsky \cite{Trushin}. Spin-polarized electron enters and exits the arc segment of the wire at $x = 0$.}
    \label{fig: wire}   
\end{center}

\begin{flalign}
H = \frac{\hbar^2 {\bf k}^2}{2m^*} + \frac{1}{2} g^* \mu_B {\bf \sigma} 
\cdot {\bf B} + H_R + H_D,                                                                                                                                                                                                                                                                                                                                                                                                                                                                                                                                                                                                                                                                                                                                                                                                        
\label{eq:hamiltonian}
\end{flalign}

\noindent where $\hbar {\bf k} = {\bf p} - e {\bf A}$ is the reduced momentum vector, $\sigma = (\sigma_x, \sigma_y, \sigma_z) \equiv (\sigma_r, \sigma_{\phi}, \sigma_z) $ are the Pauli matrices, $m^*$ the effective electron mass, $\mu_B$ the Bohr magneton and $g^*$ the effective Land\'{e} factor. For the vector potential, we use a symmetric gauge ${\bf A} = (-By/2, Bx/2, 0) \equiv (0, -Br/2, 0)$. The terms 
\vspace{-0.1cm}
\begin{flalign}
H_R = \alpha \left(  \sigma_x k_y - \sigma_y k_x \right), 
\label{eq:Rashba}\\
H_D = \beta \left( \sigma_x k_x  - \sigma_y k_y \right)
\label{eq:Dresselhaus},
\end{flalign} 

\noindent are the contributions of the Rashba and the Dresselhaus SOCs, respectively. These result from the asymmetric confining potential oriented perpendicular to the plane of the curved wire ($H_R$) and from the prevailing crystal asymmetry ($H_D$) along the growth direction (i.e. $z$-axis) of the heterostructure. Here we only consider that $H_D$ is linear in $k$; since for low-dimensional systems the Dresselhaus term proportional to $k^3$ is negligible \cite{ Berche, Schliemann} as the comparative ratio between the linear-in-k DSOC is typically much greater than the cubic one \cite{Schliemann}.  


\subsection{The curved channel}
\label{sec: curved}

The Rashba and the Dresselhaus SOC Hamiltonians, in Cartesian coordinate, are shown in Eqs.~(\ref{eq:Rashba}) and (\ref{eq:Dresselhaus}), respectively. Using the cylindrical equivalent of the momentum operators $p_x$ and $p_y$ given by
\begin{flalign}
p_x = -i\hbar \left( \cos \phi \frac{\partial}{\partial r} - \frac{\sin \phi}{r} \frac{\partial}{\partial \phi} \right)
\label{app:px}
\\
p_y = -i\hbar \left( \sin \phi \frac{\partial}{\partial r} + \frac{\cos \phi}{r} \frac{\partial}{\partial \phi} \right)
\label{app:py}
\end{flalign}
\noindent and the vector potential $\textbf{A} = (0,-Br/2,0)$, the components of the reduced momentum $\hbar \textbf{k} = \textbf{p} - e \textbf{A}$ are written as $k_x = p_x/\hbar + (\Phi/\Phi_0r) \sin \phi$ and $k_y = p_y/\hbar + (\Phi/\Phi_0r) \cos \phi$.
Here $\Phi = B\pi R^2$ is the magnetic flux through the half-ring and $\Phi_0=2\pi\hbar/e$ is the flux quantum. We substitute the above expressions to Eq.~(\ref{eq:hamiltonian}) to get the cylindrical transform of the total Hamiltonian. The resulting kinetic energy operator will yield a geometric potential arising from the local curvature in the arc segment \cite{Chaplik}. Such confinement in this section of the quantum wire implies that here, $r=R$, the potential is small and outside this region it is large. This consequently takes the average values of $\langle 1/r \rangle \rightarrow 1/R $ and $\langle \partial/ \partial r \rangle \rightarrow -1/2R$ \cite{Meijer}. Hence, for the curved segment of the wire, we conveniently write the Hamiltonian operator in cylindrical coordinates into
\begin{flalign}
H &= \frac{\hbar^2}{8m^*R^2} + \frac{\hbar^2}{2m^* R^2} \left( i \frac{\partial}{\partial \phi} + \frac{\Phi}{\Phi_0}\right)^2 + \frac{1}{2} g^* \mu_B \sigma_z B + H_R + H_D, 
\label{eq:totalHamiltonian}
\end{flalign}

\noindent where the first term takes into account the kinetic energy of the spin-polarized electron moving along the confined region, $r=R$. This assures that the above Hamiltonian is Hermitian \cite{Meijer}. The explicit cylindrical polar forms of $H_R$ and $H_D$ are, respectively, given by

\vspace{-0.1cm}
\begin{flalign}
H_R + H_D &= -\frac{\alpha}{R} \sigma_r \left( i \frac{\partial}{\partial \phi} + \frac{\Phi}{\Phi_0}\right) -\frac{i \alpha}{2R} \sigma_{\phi} - \frac{\beta}{R} \sigma_{\phi} \left( i \frac{\partial}{\partial \phi} + \frac{\Phi}{\Phi_0}\right) 
+ \frac{i \beta}{2R} \sigma_r   .
\label{app: SOCpolar}
\end{flalign}

\noindent where $\sigma_r = \sigma_x \cos \phi + \sigma_y \sin \phi$ and $\sigma_{\phi} = - \sigma_x \sin \phi + \sigma_y \cos \phi$.  Here we also note of the following relations, \ $\partial \sigma_r/\partial \phi = \sigma_{\phi}$ and $\partial \sigma_{\phi}/\partial \phi = - \sigma_r$.
 
Solution $\Psi(\phi)$ to the Schr\"{o}dinger equation with the Hamiltonian given in Eq.~(\ref{eq:totalHamiltonian}) is expressed as a linear superposition of orthogonal eigenvectors $e^{in\phi}/\sqrt{2 \pi}$ where $n$ is an integral wave number. In general, $\Psi$ is given by
\vspace{-0.1cm}
\begin{flalign}
\Psi(\phi) = \frac{1}{2 \pi} \sum_n \left( \begin{array}{c}  C_n^+ e^{in\phi} \\ C_n^- e^{in\phi} \end{array}\right)
\end{flalign}

\noindent where $C_n^+$ and $C_n^-$ are the coefficients of the spin-up and spin-down states, respectively. Solving exactly the Schr\"{o}dinger equation $H \Psi = E_{n,\phi} \Psi$ gives us the corresponding eigenvalues for the arc segment in the form
\vspace{-0.9cm} 

\begin{flalign}
E_F^s &= \frac{\varepsilon_0}{4} + \varepsilon_0 \left( {q_R} + \frac{1}{2} \right)^2 + s \left[ \left[ \varepsilon_0 \left( {q_R} + \frac{1}{2} \right) - \varepsilon_Z \right]^2 +  \frac{\left( \alpha^2 + \beta^2 \right)}{R^2} \left( q_R + \frac{1}{2} \right)^2 \right]^{1/2}  
\label{eq: arc_energies}
\end{flalign} 

\noindent for a right-moving spin with angular momentum $q_R$ where $\varepsilon_0=\hbar^2/2mR^2$. Here, the energy is fixed at the Fermi energy, $E_F$, Thus,  Eq. (\ref{eq: arc_energies}) is the energy expression for the eigenstates at the Femi level, which we consider mainly contributes in the transport. Coefficient $s = \pm 1$ refers to the spin state and $\varepsilon_Z = g^* \mu_B B/2$ is the Zeeman energy. The relation for a left-moving spin is obtained with $q_R$ replaced with $-q_L$. 

On the other hand, the corresponding eigenvectors are as follows:
\vspace{-0.1cm}
\begin{flalign}
\Psi^+_{arc} = \left( \begin{array}{c} B^+ \cos^+ \xi_R e^{i q_R^+ \phi} + C^+ \cos^+ \xi_L e^{-i q_L^+ \phi} \\
B^- \sin^+ \xi_R e^{i q^+_R \phi} e^{i \phi} - C^+ \sin^+ \xi_L e ^{-i q_L^+ \phi} e^{i \phi}   
\end{array} \right) 
\label{eq: spinup_vectors}\\
\Psi^-_{arc} = \left( \begin{array}{c} - B^- \sin^- \xi_R e^{i q_R^- \phi} + C^- \sin^- \xi_L e^{-i q_L^- \phi} \\
B^- \cos^- \xi_R e^{i q^-_R \phi} e^{i \phi} + C^- \cos^- \xi_L e ^{-i q_L^- \phi} e^{i \phi}   
\end{array} \right) 
\label{eq: spindn_vectors}
\end{flalign}

\noindent where the angles $\xi_R$ and $\xi_L$ satisfy the following relations: 
\vspace{-0.1cm}
\begin{flalign}
	\tan^\pm \xi_R &= \frac{ 1}{\frac{a \pm i\beta}{R}\overline{q}_R^{\pm}} \left( \varepsilon_0 \overline{q}_R^\pm  - \varepsilon_Z \right) + \frac{ 1}{\frac{a \pm i\beta}{R}\overline{q}_R^{\pm}} \sqrt{(\varepsilon_0 \overline{q}_R^\pm - \varepsilon_z)^2 + \frac{\nu^2}{R^2} (\overline{q}_R^\pm)^2 } 
	\label{tanR} 
	\\
	\tan^\pm \xi_L &= \frac{1}{\frac{a \pm i\beta}{R} \overline{q}_L^\pm} \left( -\varepsilon_0 
	\overline{q}_L^\pm  - \varepsilon_Z \right) + \frac{1}{\frac{a \pm i\beta}{R} \overline{q}_L^\pm}  \sqrt{(\varepsilon_0 \bar{q}_L^\pm + \varepsilon_z)^2 + \frac{\nu^2}{R^2} ( \overline{q}_L^\pm )^2 } 
	\label{tanL}
\end{flalign} 

\noindent with $\overline{q}_R^{\pm} = q_R^{\pm} + 1/2, \overline{q}_L^{\pm} = q_L^{\pm} - 1/2$ and we set $\nu=\sqrt{\alpha^2 + \beta^2}$ as the effective SOC strength. The coefficients $B$ and $C$ are the transmitted and reflected wave amplitudes which are to be determined upon imposition of the appropriate boundary conditions. 
 
\subsection{Input and output channels}
\label{sec: input}

For the input and the output straight segments, the wave function $\Psi(x)$ satisfies the one-dimensional Schr\"{o}dinger equation $H \Psi(x) = E \Psi(x)$. Using Eqs.~(\ref{eq:hamiltonian}) to (\ref{eq:Dresselhaus}), one obtains the eigensolutions in the form of $e^{ik^{\pm}x}$ in the input and output segments:
\vspace{-0.1cm}
\begin{flalign}
\Psi^+_{in}(x) = e^{i\frac{\Phi}{\Phi_0R}x} \left( \begin{array}{c} \cos \gamma^+ \left( A_0^+ e^{ik^+x} + A^+e^{-ik^+x}\right) \\
-i \sin \gamma^+ \left( A_0^+ e^{ik^+x} - A^+e^{-ik^+x}\right) 
\label{eq: inputup_vectors}
\end{array}\right)
\\
\Psi^-_{in}(x) = e^{i\frac{\Phi}{\Phi_0R}x} \left( \begin{array}{c} -i \sin \gamma^- \left( A_0^- e^{ik^-x} - A^-e^{-ik^-x}\right) \\
\cos \gamma^- \left( A_0^- e^{ik^-x} + A^-e^{-ik^-x}\right)
\end{array}\right)
\label{eq: inputdn_vectors}
\end{flalign}

\vspace{-0.1cm}
\begin{flalign}
\Psi^+_{out}(x) = e^{i\frac{\Phi}{\Phi_0R}x} \left( \begin{array}{c} \cos \gamma^+ D^+ e^{ik^+x}  \\
i \sin \gamma^+ D^+ e^{ik^+x} 
\end{array}\right)
\label{eq: outputup_vectors}
\\
\Psi^-_{out}(x) = e^{i\frac{\Phi}{\Phi_0R}x} \left( \begin{array}{c} i \sin \gamma^-  D^- e^{ik^-x}  \\
\cos \gamma^-  D^- e^{ik^-x} 
\end{array}\right)
\label{eq: outputdn_vectors}
\end{flalign}
   
\noindent where the superscript (+, -) corresponds to the electron spin state. The coefficients $A_0^{\pm}$ and $A^{\pm}$ are the incoming and reflected wave probability amplitudes, respectively, while $D^{\pm}$ is the output transmitted wave amplitude. In the assumption that the input electron is spin-polarized we set $A_0^+ =1$ and $A_0^- =0$. The dimensionless quantity $\gamma^{\pm}$ is given by
\vspace{-0.1cm}
\begin{flalign}
\tan \gamma^{\pm} = \frac{-\varepsilon_Z + \sqrt{ \nu^2 (k^{\pm})^2 + \varepsilon_Z^2} }{(\alpha \mp i\beta) k^{\pm}}.
\end{flalign}  

\noindent The eigenvalues, in this specified region, take the form 
\vspace{-0.1cm}
\begin{flalign}
E^{\pm} = \frac{\hbar^2 (k^{\pm})^2}{2m^*} \pm \sqrt{\nu^2(k^{\pm})^2 + \varepsilon_Z^2}
\end{flalign}

\noindent from which the energy gap $\Delta E = |E^+ - E^-|$ is found to be dependent on the magnetic field and on the SOCs.

All of the unknown coefficients are determined by imposing continuity of the eigenvectors at the boundary between the curved 1D wire and the input or output channel as follows: 
\vspace{-0.3cm}

\begin{flalign}
&\Psi_{in}(x) = \sum\limits_{\mu=+,-} \Psi_{in}^{\mu}(x)
\qquad &x \le 0,  \\
&\Psi_{arc}(\phi) = \sum\limits_{\mu=+,-} \Psi_{arc}^{\mu}(\phi) \qquad &\phi \subseteq [-\pi/2, \pi/2],
\\
&\Psi_{out}(x) = \sum\limits_{\mu+,-} \Psi_{out}^{\mu}(x) \qquad &x \le 0,
\end{flalign}

\noindent including their first derivatives. From here we can relate the wave amplitudes of the eigenfunctions in the three segments.

\subsection{Transport Properties}

In this work, we set that the electron is in an initial spin up state ($A_0^+=1$, $A^-_0=0$) with a corresponding input polarization $P_{in}=1$. 

At the output, we obtain the spin probability current density from the continuity equation as follows 

\begin{flalign}
j^{\pm}_{out} &= \frac{\hbar}{2m^*} \left( \Psi^{\pm}_1 \hat{k}^*_x \Psi^{\pm *}_1 + \Psi_1^{\pm *} \hat{k}_x \Psi_1 + \Psi_2 \hat{k}^*_x \Psi^*_2 + \Psi_2^* \hat{k}_x \Psi_2\right) + \frac{1}{\hbar} \left[ \left( \beta - i\alpha \right) \Psi_1 \Psi_2^* + \left( \beta + i\alpha \right) \Psi^*_1 \Psi_2\right] 
\label{eq: output_current1}
\end{flalign}   

\noindent where $\Psi^{\pm}_1$ and $\Psi^{\pm}_2$ are obtained from Eqs.~(\ref{eq: outputup_vectors}) and (\ref{eq: outputdn_vectors}) when we express $\Psi^{\pm}_{out} = \left(\begin{array}{c} \Psi_1^{\pm} \\ \Psi_2^{\pm} \end{array} \right) $. Evaluating the above equation, we arrive at  
\vspace{-0.1cm}
\begin{flalign}
j_{out}^{\pm} = |D^{\pm}|^2 \frac{\hbar k^{\pm}}{m^*} \left[ \nu^2 (k^{\pm})^2 + \sin^2 \gamma^{\pm} \pm \frac{2m^*}{\hbar^2} \nu^2 \sin \gamma^{\pm} \right],
\label{eq: output_current}
\end{flalign}

Spin switching, which is characterized by flipping the initial spin state of a particle into its opposite final state, is seen from the calculation of the output spin polarization that is given by  $P_{out}= (j^+_{out} - j^-_{out})/(j^+_{out} + j^-_{out})$. In Eq.~(\ref{eq: output_current}) we disregard the effect of temperature. Since our system does not account for any electron-electron interaction, the temperature dependence of the transport properties comes in with the consideration of the Fermi function as done in Ref.~\cite{Molnar}. 

Another transport property of interest is the spin-dependent  conductance. At the output, this is proportional to the probability density $|D^{\pm}|$ as 
\vspace{-0.1cm}
\begin{equation}
G^{\pm} = \frac{e^2}{2 \pi \hbar} |D^{\pm}|^2, 
\label{eq: conductance}
\end{equation}

\noindent with the total spin conductance of the quantum wire given by $G = G^+ + G^-$. Reversal of the direction of the magnetic field can be observed through the
magnetoconductance ratio (MC) defined as 

\vspace{-0.3cm}
\begin{equation}
MC = \left| \frac{G - G^*}{G + G^*} \right|
\end{equation}
 
\noindent where $G^*$ is the value of the total spin conductance when the magnetic field direction is reversed. The value of MC is a measure of  the effectiveness of the external magnetic field in spin switching.

All transport properties mentioned such as spin probability current densities, spin polarization and spin conductance involve the determination of the transmission coefficients $D^{\pm}$, which is carried out numerically using transfer-matrix approach in Mathematica 9.0.  Here, relations between the wave amplitudes of the segments are summarized in matrix form as follows:

\begin{flalign}
&M_{inc} \left[ \begin{array}{c} A_0^+ \\ A^+ \\ A_0^- \\ A^- \end{array} \right] =  M_{arc,1} \left[ \begin{array}{c} B^+ \\ C^+ \\ B^- \\ C^- \end{array} \right] \\
&M_{arc,2} \left[ \begin{array}{c} B^+ \\ C^+ \\ B^- \\ C^- \end{array} \right] =  M_{out} \left[ \begin{array}{c} D^+ \\ D^- \end{array} \right]
\end{flalign}  

\noindent where $M_{inc}$ relates the transmission and the reflection coefficients, $A_0^{\pm}$ and $A^{\pm}$ in the input segment to the wave amplitudes in the curved segment and $M_{out}$ links the wave amplitudes in the curved segment to the transmission coefficients  at the output. $M_{arc, \mu}$ is the corresponding matrix in the curved segment at the input ($\mu=1$) and at the output ($\mu=2$) boundaries.  The elements of these transfer matrices are obtained upon the imposition of the boundary conditions.  

\section{Numerical Results and Discussion}
\label{sec:Results}

Unlike $\alpha$ which can be externally controlled by a gate bias in semiconductor devices, the strength of the DSOC, $\beta$, is usually assumed constant as it is already pre-set during the growth of the semiconductor layers and in the fabrication of the device. Hence, in our paper, the studied transport properties are derived for a constant $\beta$. 

\begin{center}[t]
    \includegraphics[width=0.50\textwidth]{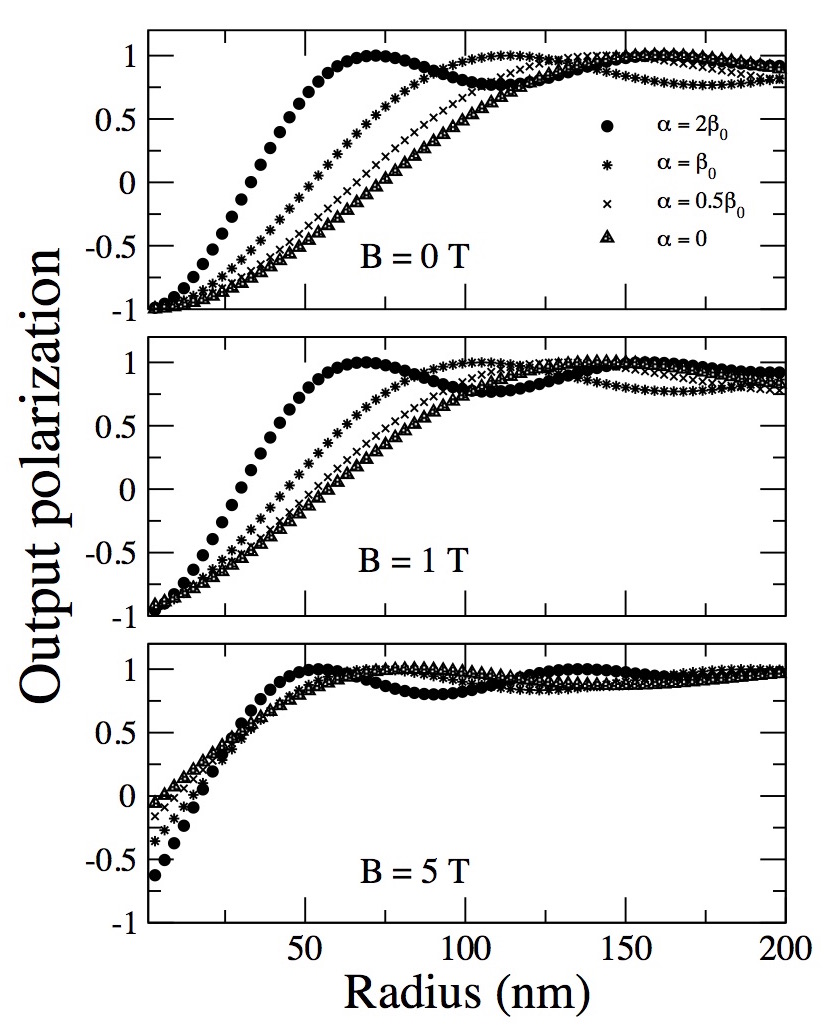}
   \captionof{figure}{Output polarization versus radius of curvature for different RSOC strength $\alpha$. The  magnetic field is varied with values: $B=0$ T, $B=1$ T and $B=5$ T.}
    \label{fig: Pola_DSOCvary}   
\end{center}

In our numerical calculations, we used the InAs parameters utilized in Ref.~\cite{Trushin}: $m^*=0.033m_e$, $g^*=-12$ and $E_F=30$~meV. We investigated the output polarization and the transport properties for the quantum wire set up with an arc radius ranging from $1$~nm to $250$~nm. This implies that the input- and output- parallel wires would be separated by a distance of only 2$R$. With the recent experiment of Ron and Dagan \cite{Ron} being able to create parallel 1D wires with only an alternating distance of one- and three- LaAlO$_3$ unit cells apart, we believe that present technologies would be able to create such a device with 2$R$ possibly below $25$~nm. From Figs.~\ref{fig: Pola_DSOCvary} to \ref{fig: MCR2D0}, a constant DSOC strength $\beta_0 = 12.48 $ meV nm is chosen within the range of SOC strength that was experimentally obtained for InAs as reported in Ref. \cite{Yang, Matsuyama, Grundler}. To determine the effectiveness of  the SOCs, the value of $\alpha$ is varied relative to $\beta$.

In Fig.~\ref{fig: Pola_DSOCvary}, the manifestation of spin switching is observed in the polarization curves for different $R$, at which the values of $P$ switches from its initial value $P_{in}=+1$ to an output polarization $P_{out} < 0$ for small $R$. This is attributed to a strong confinement energy $\varepsilon_0 >> \nu/R$ which led to a non-adiabatic transport of the spin-polarized  electron. Our result is also consistent to those previously reported in Ref.~\cite{Trushin} on the observed current density redistribution wherein only the presence of RSOC was considered. In our work, however, we found that this behavior of $P$ can still be observed even with a non-zero DSOC. In fact, the exact polarization curves shown in Fig.~\ref{fig: Pola_DSOCvary} can be achieved when we held $\alpha$ constant instead, and $\beta$ is set to a different value keeping the same ratio of $\alpha/\beta_0$ in Fig.~\ref{fig: Pola_DSOCvary}.

\begin{center}
    \includegraphics[width=0.50\textwidth]{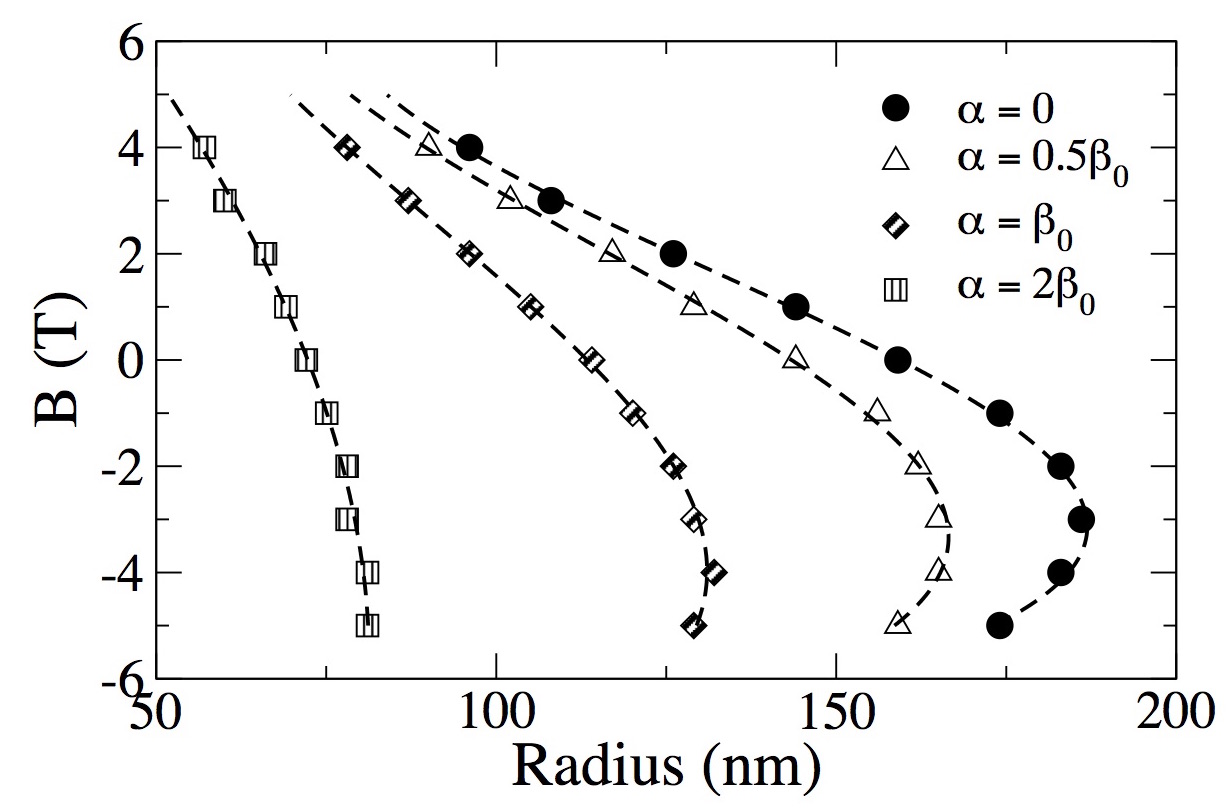}
   \captionof{figure}{Magnetic field versus radius of curvature at the time when $P_{out}$ first reaches $P_{in} = +1$.  The dashed lines represent the  $B \approx R^{1/3}$ fitting curves that are used to describe the plotted data.}
    \label{fig: BvsRad}   
\end{center}

Still in Fig.~\ref{fig: Pola_DSOCvary}, for the case when $B = 0$~T or $1$~T, an increasing $\alpha$ causes a rapid saturation in the polarization curves towards $P_{in}=+1$. This observation again follows from the results in Ref. \cite{Trushin} when $\beta = 0$. However, here we also found that this behavior of $P_{out}$ occurs even when $\alpha$ is fixed and $\beta$ is set to higher value. Thus we are led to generalize that it is the increasing effective SOC strength $\nu$ that causes the early saturation of $P_{out}$ towards $P_{in}$. The increasing influence of either RSOC and DSOC or both SOCs contribute to a larger effective SOC strength that narrows the range of $R$ at which spin switching occurs. In this case, regardless of its origin, the SOCs widen the energy gap between the spin up and the spin down states, which in turn, reduces the probability that an initially polarized electron occupies the opposite spin state. 

Now this time if we look at the effect of $B$ in $P$ in Fig.~\ref{fig: Pola_DSOCvary}, we notice that for $B=0$~T and $1$~T, $100\%$ spin switching is achieved with $P_{out} \rightarrow -1$. For $B=5$~T, however, maximum spin switching can hardly be attained even at very small radius ($R < 25$~nm), with $P_{out}$ not reaching -1. One can attribute this to a strong Zeeman effect as we go back to Eq.~(\ref{eq:hamiltonian}), which makes the electron spin to align itself to the field. Larger SOC, therefore, is necessary to perform spin modulation and to achieve optimum spin switching.  As illustrated in Fig.~\ref{fig: Pola_DSOCvary}, there is a largest change in polarization from the initial value if the effective SOC strength is greatest. 

\begin{center}
    \includegraphics[width=0.50\textwidth]{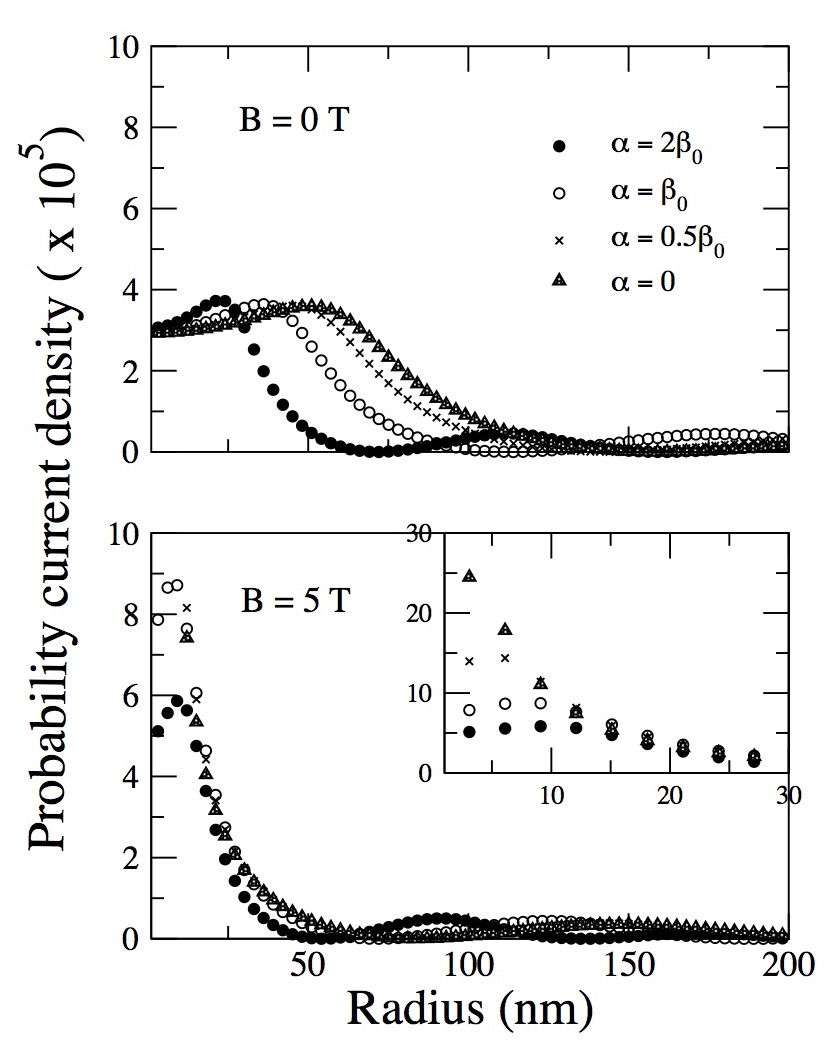}
   \captionof{figure}{Probability current density (in SI unit) at the output with spin down orientation for different RSOC strengths with $B = 0$~T and $B = 5$~T.  }
    \label{fig: Current_RSOCvary}   
\end{center}

Next, we need to have a measure of the region of the confinement radius where the spin switching occurs relative to the tunable parameters such as $\alpha$ and $B$. From Fig.~\ref{fig: Pola_DSOCvary}, we plot the value of $R$ and its corresponding $B$ at which $P_{out}$ first reaches $P_{in}=+1$ for different SOC strengths. Here, in Fig.~\ref{fig: BvsRad}, we have a clearer picture that the larger the effective SOC strength the smaller $R$ value is required for spin switching to occur. The curves here indicate the characteristic $R$ below which $P_{out} < +1$.    Through curve fitting, we are able to establish that the data, regardless of the value of $\nu$, can be described by the proportionality relation, $B \approx R^{1/3}$. This relation sets a limit to spin switching through a curved quantum wire.  

Shown in Fig.~\ref{fig: Current_RSOCvary} is the behavior of the spin down probability current density at the output channel. For the $B=0$~T case, and as $\alpha$ goes to zero, we observe a wider range of $R$ where there is large probability current density, which sinusoidally decreases as $R$ is further increased. It can be seen that the amplitude of the peaks does not differ significantly with varying SOC strength. On the other hand, for the case of $B=5$~T the amplitude of the probability current density increases due to the Zeeman contribution to the energy which is consistent  with that of Fig.~\ref{fig: Pola_DSOCvary}. The range at which there is large probability current density occurs within a region where $R<25$~nm for all $\alpha$ values. In this region, we find that its probability current density amplitude, at a given $R$, can be amplified by tuning the RSOC strength, while the DSOC strength is held constant. In the case of Fig.~\ref{fig: Current_RSOCvary} where the chosen DSOC strength is $\beta_0$, the probability current density amplitude decreases as $\alpha$ increases indicating that the effect of spin-orbit coupling is to suppress the spin down state. This cannot be easily seen in Eq.~(\ref{eq: output_current}) since the dependence on $|D^{\pm}|$ to SOC is not explicit. 

\begin{center}
    \includegraphics[width=0.60\textwidth]{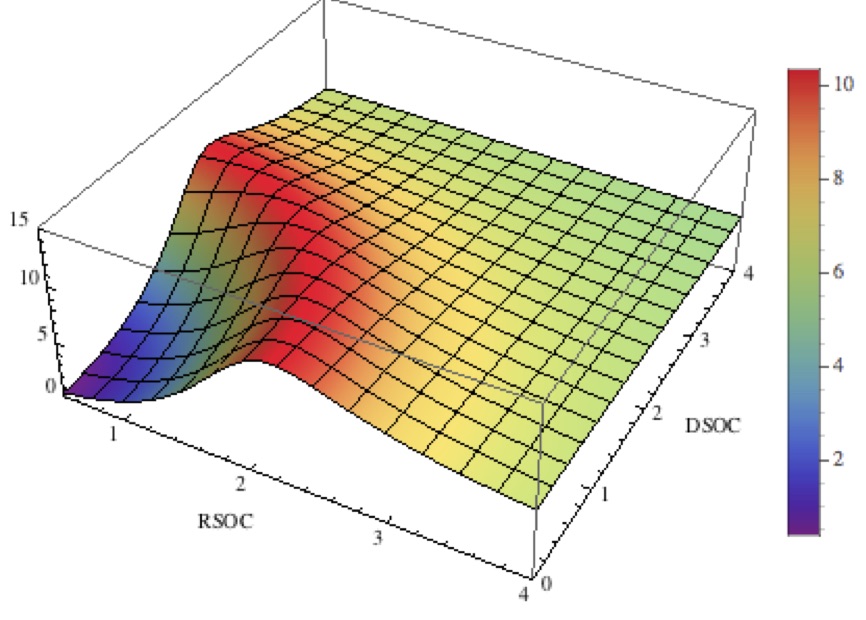}
   \captionof{figure}{(Color online) Probability current density at the output with spin down orientation for different RSOC and DSOC strengths with $R = 10$~nm and $B = 5$~T.}
    \label{fig: 3DJM10nmB5}   
\end{center}

The observed bumps (or peaks) in the current density plots in Fig.~\ref{fig: Current_RSOCvary} are attributed to the modulating effect of the RSOC strength at a particular $R$ value. As shown in Fig.~\ref{fig: 3DJM10nmB5}, for $R = 10$~nm and $B = 5$~T, we will see that there are select pairs of values of $\alpha$ and $\beta$, forming an arc, at which the current density is maximum. The radius of the arc satisfies a circle equation which follows from the earlier definition of the effective SOC strength $\nu$. Thus the selection of $\alpha$ and $\beta$ which will give large current density strongly depends on $\nu$. Since the DSOC strength is constant in a  device, a simulation such as in Fig.~\ref{fig: 3DJM10nmB5} will provide the optimal combination of $\alpha$ and $\beta$ that amplifies the current. It turns out that for high magnetic fields, the modulation effect by SOCs can occur within narrow range of $R$.       

\begin{center}
    \includegraphics[width=0.50\textwidth]{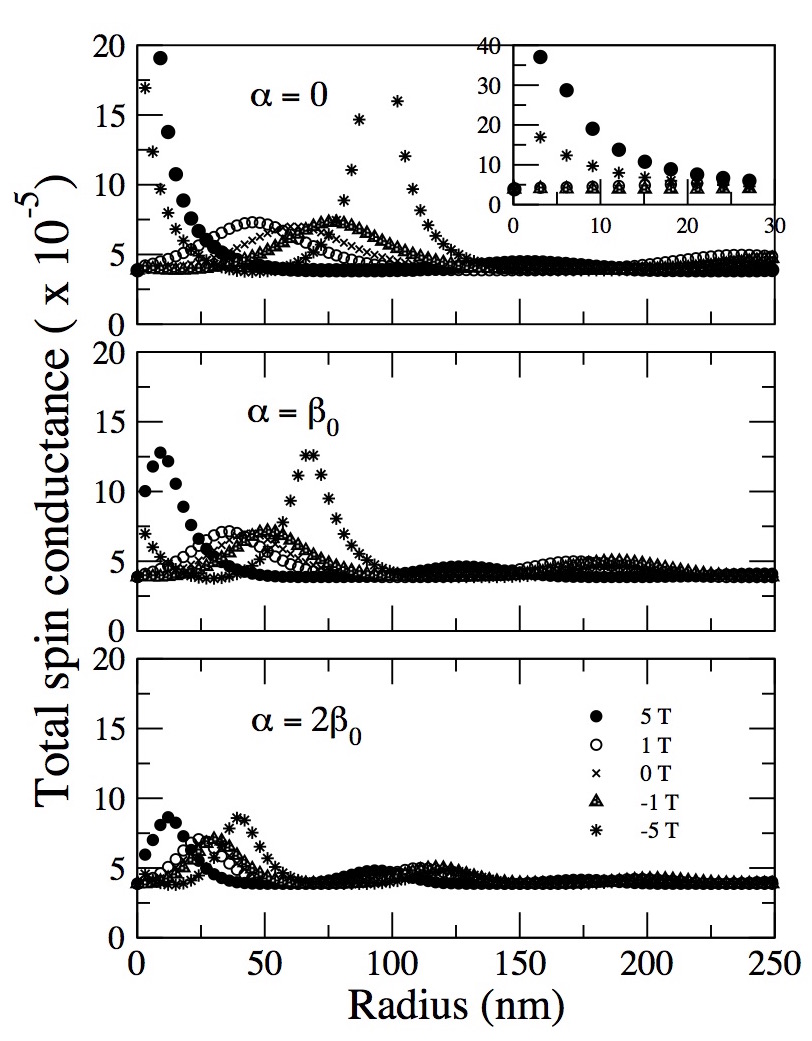}
   \captionof{figure}{Total spin conductance (in SI unit) for different magnetic field strengths with SOC strengths: $\alpha = 0$, $\alpha = \beta_0$ and $\alpha = 2\beta_0 $. }
    \label{fig: Magneto_Bvary}   
\end{center}

Fig.~\ref{fig: Magneto_Bvary} shows the variation of the total spin conductance for different $\alpha$, $R$ and $B$ for which we set $\beta_0$. Achieved results are found to be consistent with those earlier observed in the output spin polarization and in the probability current density. These include the leftward shift in the conductance curve for weak magnetic fields, as well as the insignificant difference on the amplitude, when we vary the SOC strength. 
For high fields, the system exhibits large spin conductance peak when the SOC is weak. It can be seen that reversing the direction of the applied field yields the same conductance peak amplitude but is shifted to some other value of $R$. This shift with respect to $R$ by magnetic field reversal is consistent in the plot of output polarization. 

Fig. \ref{fig: MCR2D0} shows the behavior of magnetoconductance as a function of magnetic field strengths with radius of curvature: $R = 30$~nm and $R = 300$~nm for the case when only one type of SOC is present, $\alpha = 0$, and when both RSOC and DSOC of equal strengths are present, $\alpha = \beta_0$. Here we find that a high MC ratio reaching 40\% can be obtained for $R = 30$~nm. For large $R$, however, MC decreases significantly.  On the other hand, the presence of both types of SOCs increases the effective SOC strength in the system which leads to a decrease in the MC. This result reaffirms the previous observation on the reduced effect of $B$ on spin switching when SOC is large.  

\begin{center}
    \includegraphics[width=0.50\textwidth]{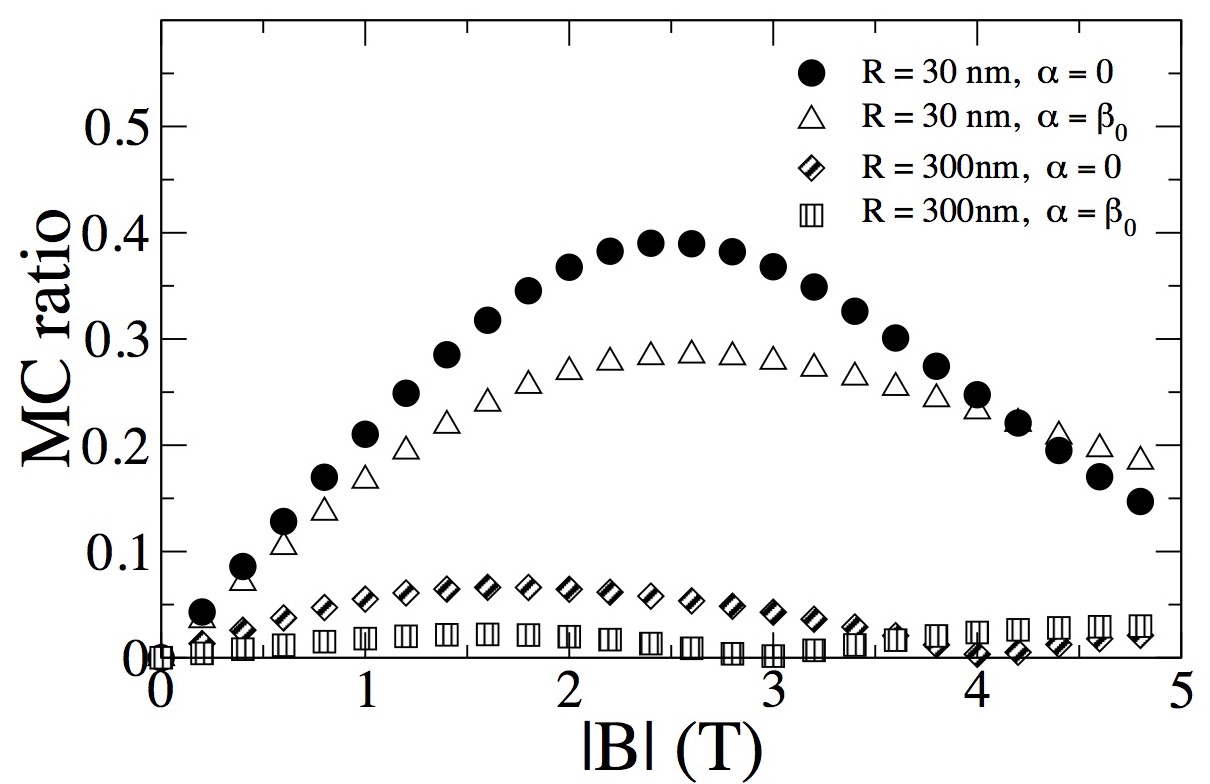}
    \captionof{figure}{Magnetoconductance (MC) ratio as a function of magnetic field strength for $R = 30$~nm and $R = 300$~nm and for $\alpha = 0$ and $\alpha = \beta_0$.  }
    \label{fig: MCR2D0}   
\end{center}
 
The results in this work applies to devices with semicircular quantum wires of fixed radius of curvature. When the radius is no longer fixed, the first term of the Hamiltonian in Eq.~(\ref{eq:totalHamiltonian}) is no longer applicable. The confinement condition should be changed accordingly. For other forms of curved wire structures, such as the two-dimensional curved waveguides in Ref.~\cite{Bulkagov}, the method of calculating the energies and the transport properties presented in this paper will essentially hold. However, several considerations must be accounted for -- such as the changes in the appropriate boundary conditions between the segments, the existence of different transmission modes and intersubband mixing, as well as, the cubic Dresselhaus spin-orbit coupling contribution. The latter was found to provide more of the spin-flipping rather than the linear $k$ in 2D electron quantum dots \cite{Krich}.
\vspace{-0.25cm} 

\section{Conclusion}
\label{sec: Conclusion}
In our work, we have derived the eigenvalues and eigenfunctions for a curved quantum wire with a confinement radius $R$ in a magnetic field in the presence of Rashba and Dresselhaus spin orbit couplings and the Zeeman interaction. For a set condition of $R$ and DSOC, tuning the RSOC and $B$ is necessary to find the best combination of $\alpha$ and $\beta$ that will give us optimal values for the spin polarization, probability current density and total spin conductance. We find that the influence of the magnetic field are found to be suppressed when the prevailing effective SOC strength is strong, as observed in the polarization curves for a given $R$. Electron spin transport in a curved wire system is found to exhibit high MC ratio up to 40\% for $R = 30$~nm and $B \approx 2.5$~T. A characteristic large spin polarization and high magnetoconductance are essential requisites in the development of spintronic devices.

\vspace{0.5cm}

\noindent \textbf{Acknowledgment}
\vspace{0.5cm}

This work is supported through the Ph.D. scholarship given to C. Baldo by the Commission on Higher Education through the National Institute of Physics as a Center of Excellence Program.





\bibliographystyle{elsarticle-num}
\bibliography{<your-bib-database>}

\begin{thebibliography}{100}

\bibitem{Wolf} S.~Wolf, D.~Awschalom, R.~Buhrman, J.~Daughton, S.~von Molnar, M.~Roukes. A.~Chtchelkanova, D.~Treger, Science \textbf{294} (2001) pp.~1488-1494

\bibitem{Rashba} E.I.  Rashba, Physica E \textbf{20} (2004) pp.~189-195

\bibitem{Lopez} C.~Lopez-Bastidas, J.~Maytorena, F.~Mireles, Phys. Stat. Sol C \textbf{4} 11 (2007) pp.~4229-4235

\bibitem{Nakhmedov} E.P.~Nakhmedov, O.~Alekperov, Eur. Phys. J. B \textbf{85} (2012) 298

\bibitem{Xu} W.~Xu, Y.~Guo, Phys. Lett. A \textbf{340} (2005) pp.~281-289

\bibitem{Lu} J.-D.~Lu, Y.-B.~Li, Superlattices and Microstructures \textbf{48} (2010) pp.~517-522

\bibitem{Xiao} Y.-C.~Xiao, W.-Y.~Ding, W.-J.~Deng, R.~Zhu, R.-Q.~Wang, Phys. Lett. A \textbf{377} (2013) pp.~817-821 

\bibitem{Jiang} Y.~Jiang, M.B.A.~Jalil, J. Phys.: Condens. Matter \textbf{15} (2003) L31

\bibitem{Trushin} M.~Trushin, A.~Chudnovksy, JETP Letters \textbf{83} 8 (2006) pp.~318-322

\bibitem{Wang} X.F.~Wang, P.~Vasilopoulos, Phys. Rev B \textbf{72} (2005) 165336

\bibitem{Wang2} X.F.~Wang, P.~Vasilopoulos, Physica E,  \textbf{39} (2007) pp.~159-165

\bibitem{Molnar} B.~Moln\'{a}r, F.M.~Peeters, P.~Vasilopoulos, Phys. Rev. B \textbf{69} (2004) 155335

\bibitem{Sil} S.~Sil, S.K.~Maiti and A.~Chakrabarti,  J. Appl. Phys. \textbf{112} (2012) 024321

\bibitem{Liu} D.-Y.~Liu, J.-B.~Xia, J. Appl. Phys. \textbf{115} (2014) 044313

\bibitem{Yang} C.H.~Yang, M.J.~Yang, K.A.~Cheng, J.C.~Culbertson, Physica E \textbf{17} (2003) 161-163

\bibitem{Ron} A.~Ron, Y.~Dagan, Phys. Rev. Lett. \textbf{112} (2014) 136801

\bibitem{Yang2} C.H.~Yang, M.J.~Yang, K.A.~Cheng, J.C.~Culbertson, Phys. Rev. B \textbf{66} (2002) 115306

\bibitem{Bulkagov} E.N.~Bulkagov, A.F.~Sadreev, Phys. Rev. B \textbf{66} (2002) 075331 

\bibitem{Berche} B.~Berche, C.~Chatelain, E.~Medina, Eur. J. Phys. \textbf{31} (2010) pp.~1267-1286

\bibitem{Schliemann} J.~Schliemann, D.~Loss, Phys. Rev. B \textbf{68} (2003) 165311 

\bibitem{Chaplik} A.V.~Chaplik, R.H.~Blick, New J Phys. \textbf{6} (2004) 33

\bibitem{Meijer} F.~Meijer, A.F.~Morpurgo, T.M.~Klapwijk, Phys. Rev. B \textbf{66} (2002) 033107.

\bibitem{Matsuyama} T.~Matsuyama, R.~K\"{u}rsten, C.~Mei$\beta$ner, U.~Merkt, Phys. Rev. B \textbf{61} (2000) 15588

\bibitem{Grundler} D.~Grundler, Phys. Rev. Lett \textbf{84} (2000) 6074

\bibitem{Krich} J.J.~Krich, B.I.~Halperin, Phys. Rev. Lett. \textbf{98} (2007) 226802
  
\end{thebibliography}



\end{document}